\documentclass{iau} 
\usepackage{graphicx}

\title[Extinction in young massive clusters] 
{Do you know the extinction in your young massive cluster?}

\author[De Marchi, Panagia \& Sabbi]   
{Guido De Marchi$^1$, Nino Panagia$^2$, Elena Sabbi$^2$}

\affiliation{$^1$ESTEC, 2200 AG Noordwijk, The Netherlands,  email: 
{\tt gdemarchi@esa.int} \\[\affilskip]
$^2$STScI, Baltimore, MD 21218, United States, email: {\tt panagia@stsci.edu, sabbi@stsci.edu}}

\pubyear{2015}
\volume{316}  
\jname{Formation, evolution, and survival of massive star clusters}
\editors{A. Nota \& C. Charbonnel, eds.}
\begin{document}

\maketitle


Up to ages of $\sim 100$\,Myr, massive clusters are still swamped in large
amounts of gas and dust, with considerable and uneven levels of extinction.
At the same time, large grains (ices?) produced by type II supernovae
profoundly alter the interstellar medium (ISM), thus resulting in extinction
properties very different from those of the diffuse ISM. To obtain 
physically meaningful parameters of stars, from basic luminosities and 
effective temperatures to masses and ages, we must understand and measure 
the local extinction law. This problem affects {\em all} the massive young 
clusters discussed in his volume.

We have developed a powerful method to unambiguously determine the 
extinction law in an uniform way across a cluster field, using multi-band 
photometry of red giant stars belonging to the red clump (RC). In the 
Large Magellanic Cloud (LMC), with about 20 RC stars per arcmin$^2$, we 
can easily derive a solid and self-consistent absolute extinction curve 
over the entire wavelength range of the photometry. Here, we present the 
extinction law of the Tarantula nebula (30 Dor) based on thousands 
of stars observed as part of the Hubble Tarantula Treasury Project 
(HTTP; Sabbi et al. 2013, 2015). Space limitations only allow us to 
highlight the main results. For more details see De Marchi et al. (2015).

\begin{figure}[b]
\vspace*{-0.5cm}
\centering
 \resizebox{0.95\hsize}{!}{\includegraphics{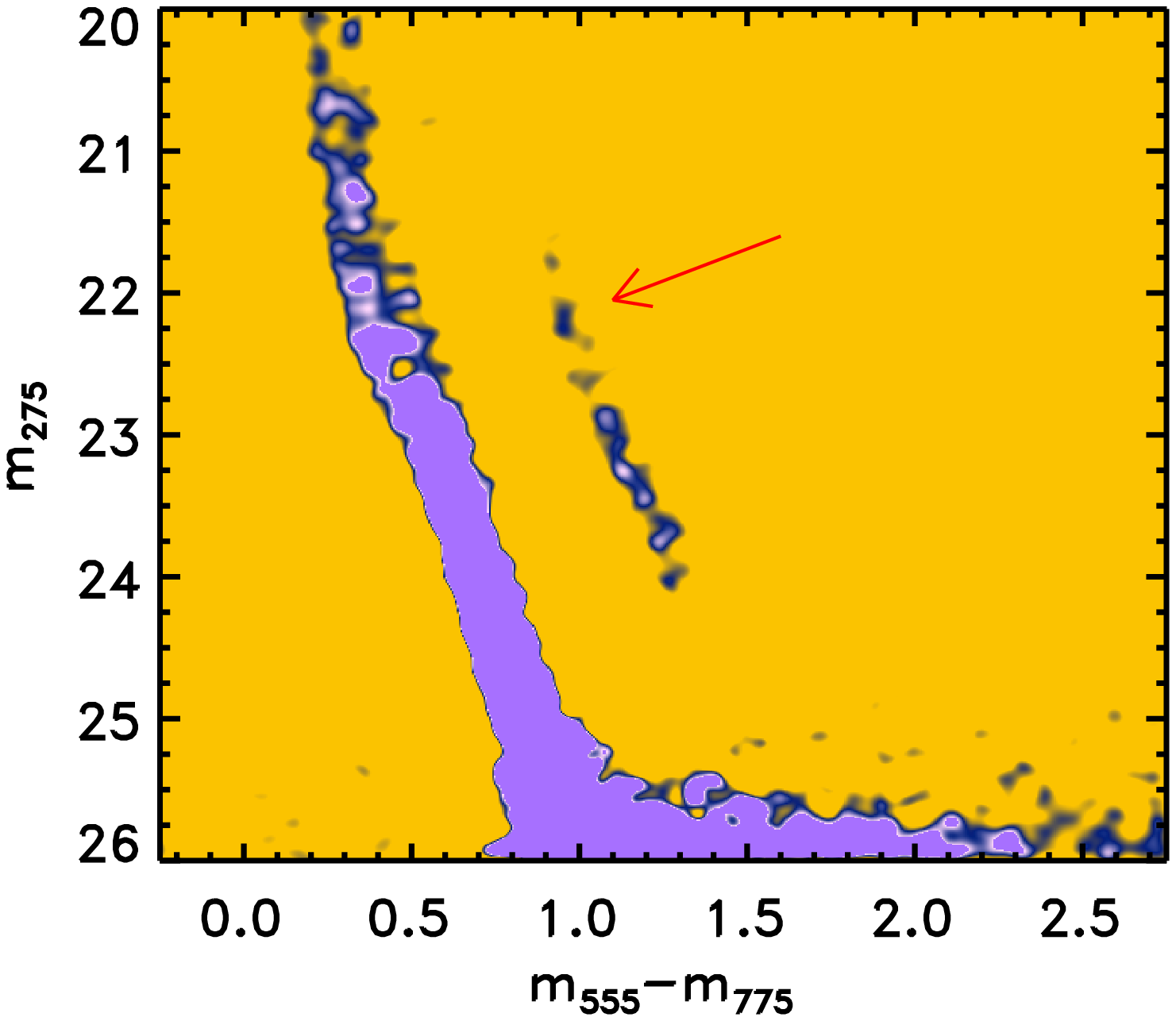}
                       \includegraphics{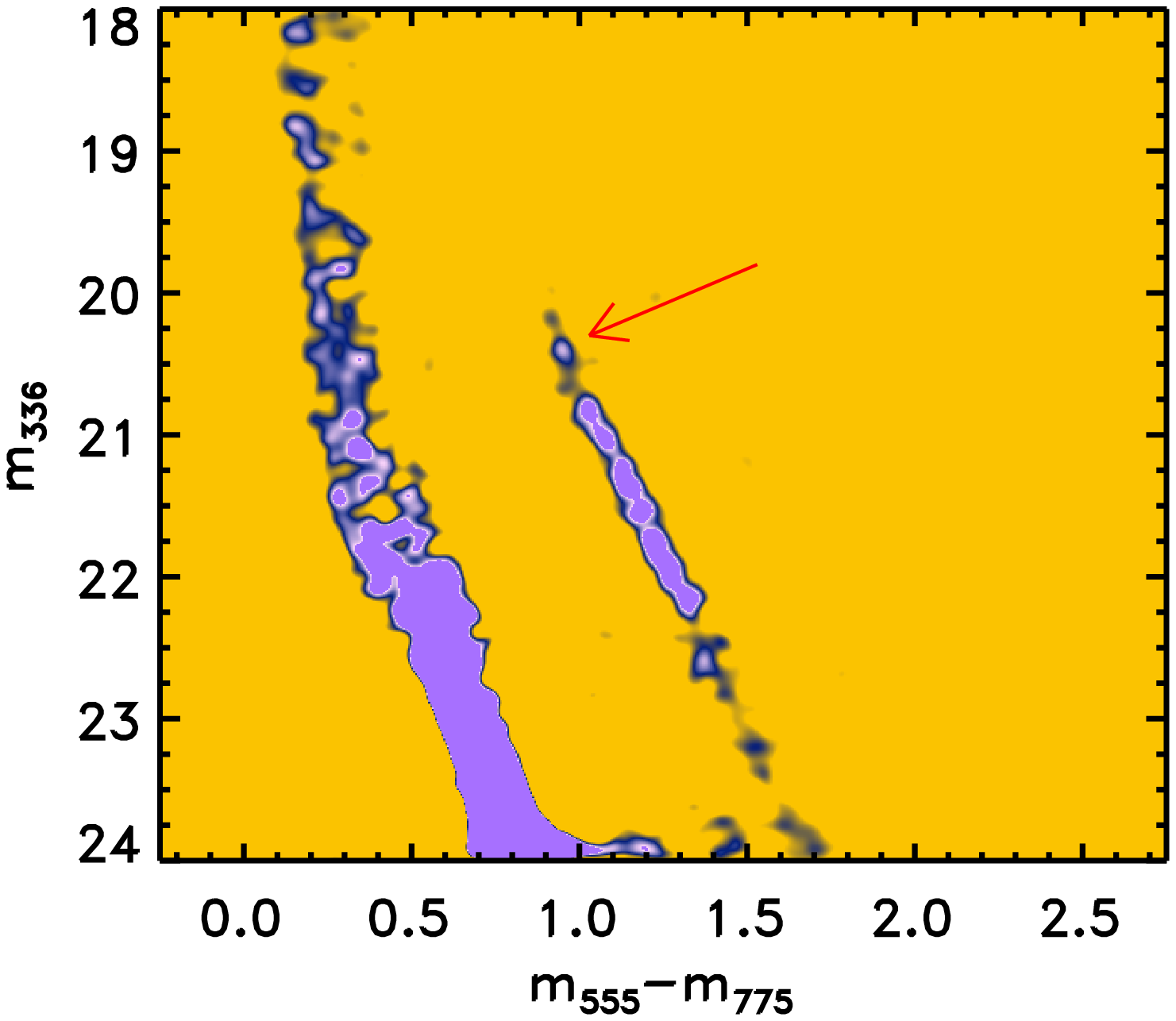}
                       \includegraphics{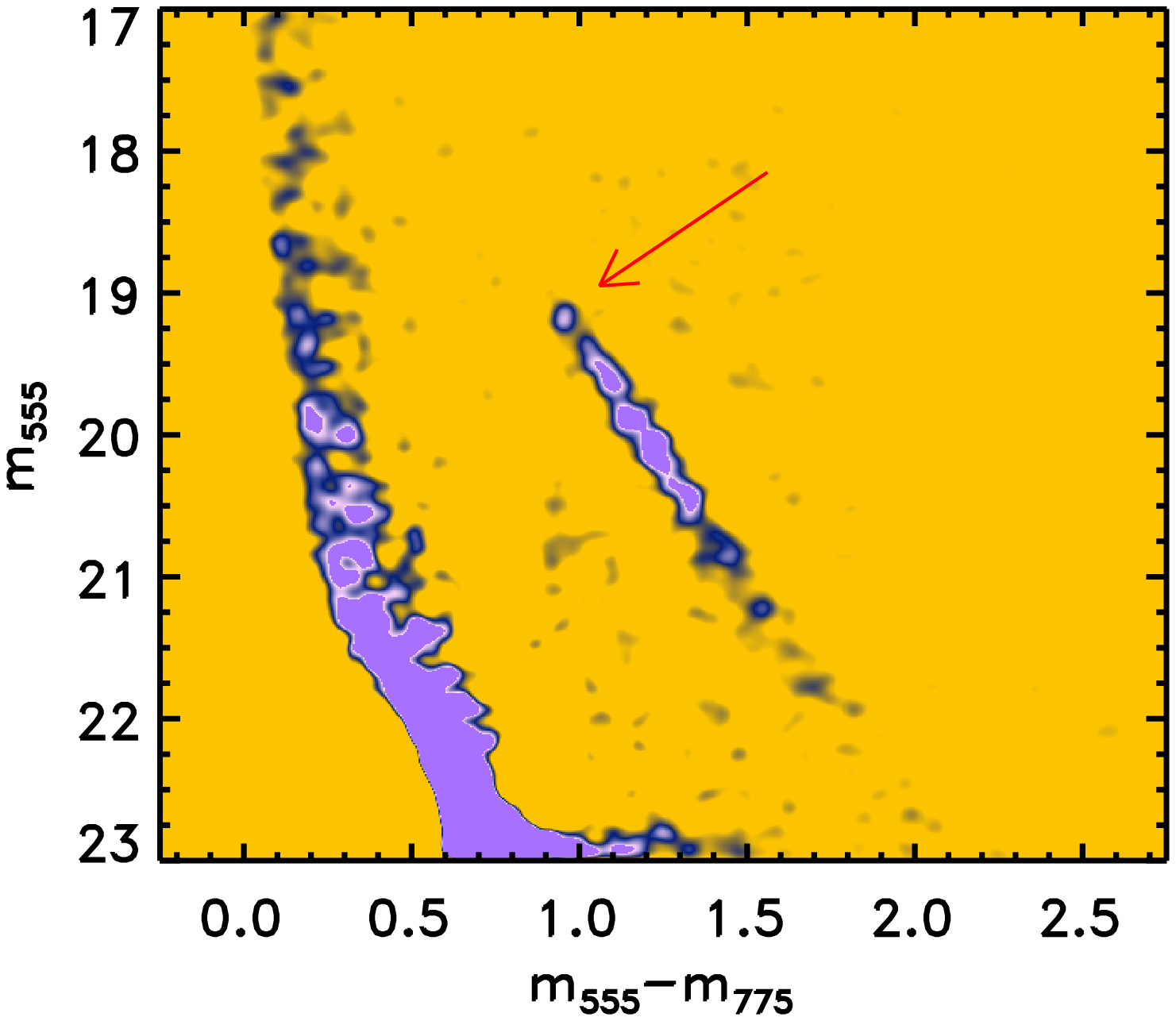}}
 \resizebox{0.95\hsize}{!}{\includegraphics{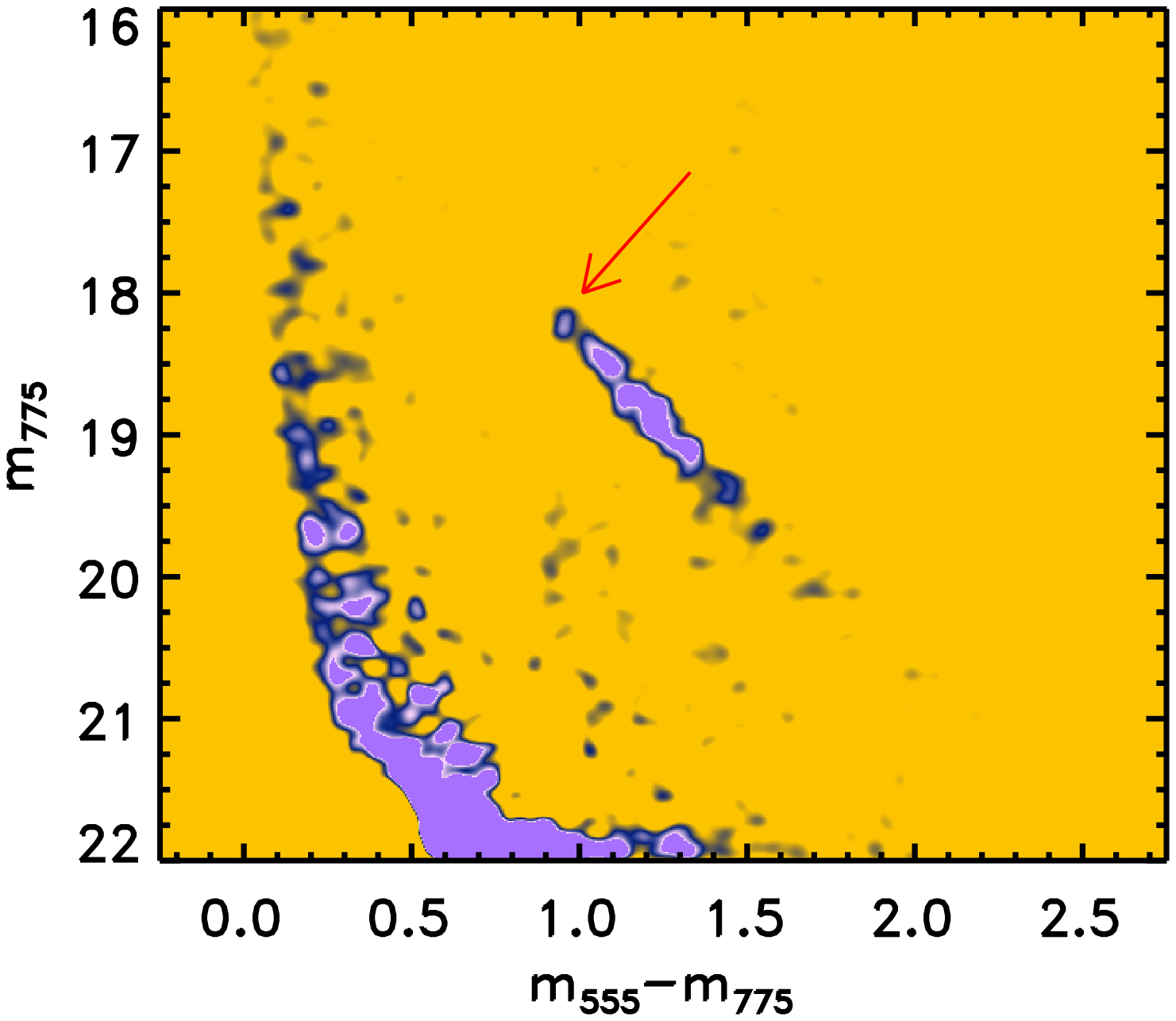}
                       \includegraphics{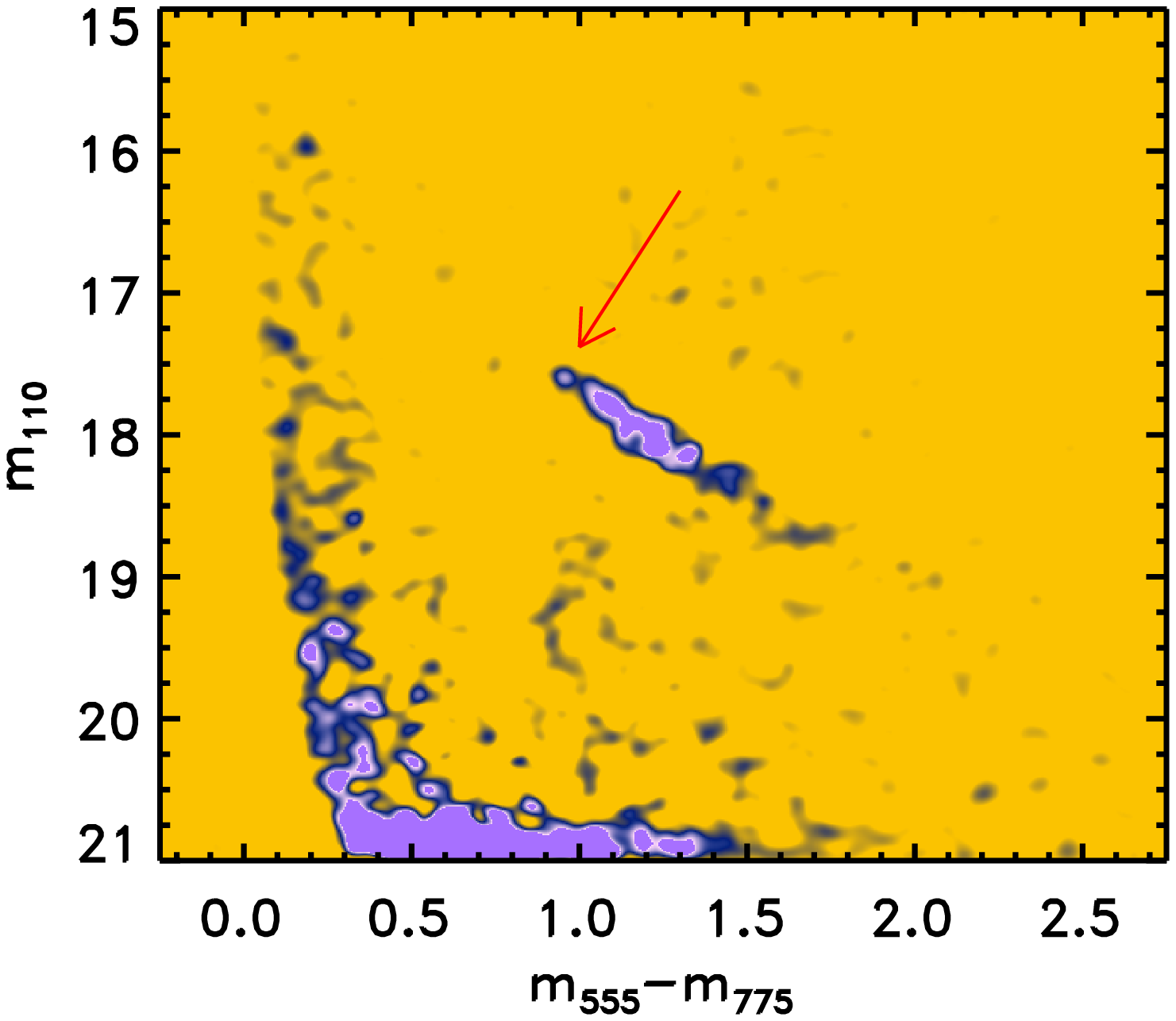}
                       \includegraphics{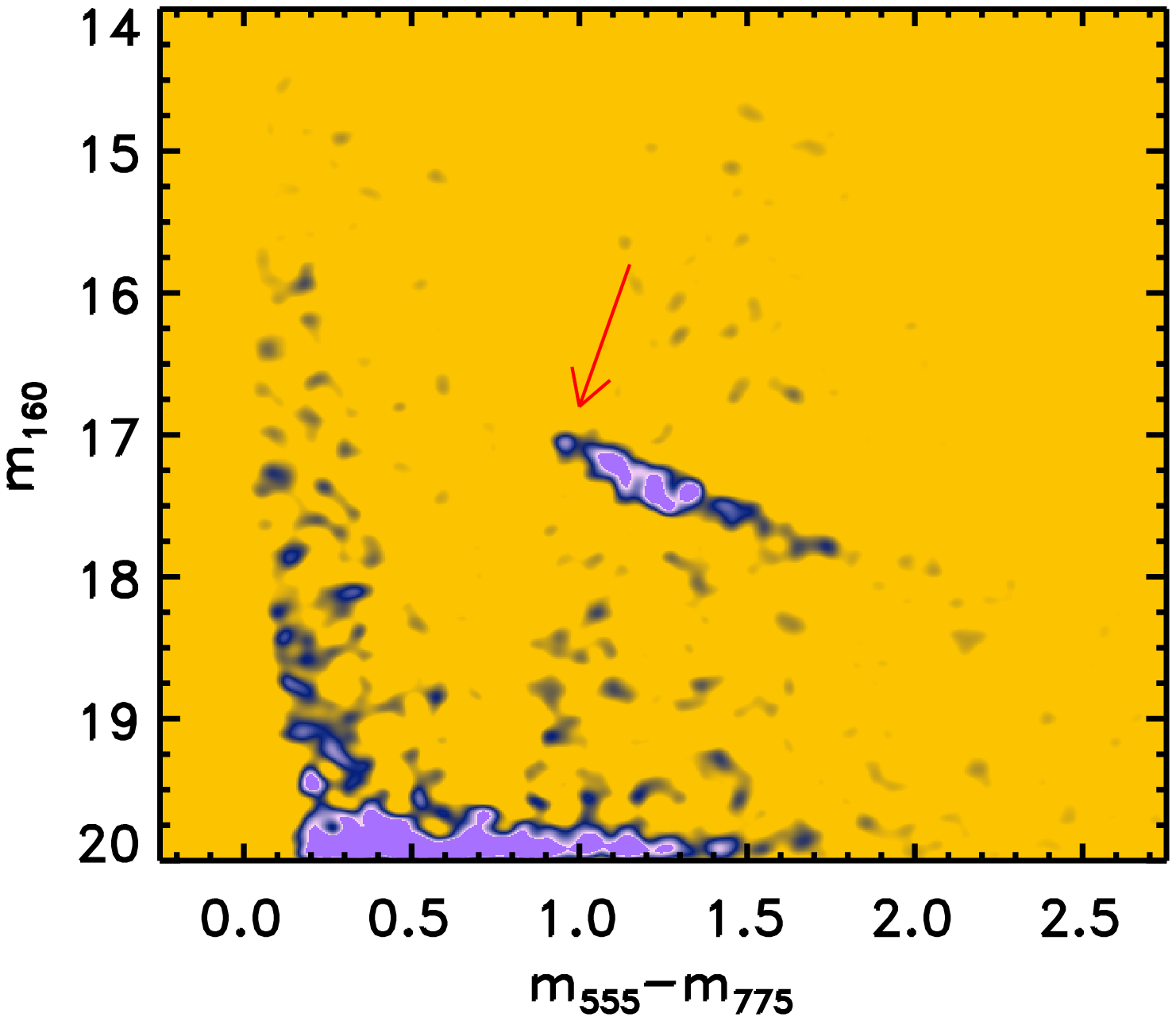}}
\caption{Unsharp masking applied to all CMDs clearly reveals the
location of the un-extinguished RC. The reddening vector is obtained 
through a linear fit to the elongated RC.}
\label{fig1}
\end{figure}

From the colour-magnitude diagrams (CMD) we can derive the absolute 
extinction in all bands towards $\sim 3\,500$  red giants that belong to 
the RC. The arrows in Fig.\,\ref{fig1} indicate the position of the 
un-reddened RC. The high and variable levels of extinction spread RC 
stars across the CMD defining tight strips. The slopes of such 
strips in all bands provide a direct measure of the reddening vector. 

\begin{figure}
\vspace*{-0.5cm}
 \begin{minipage}[c]{0.67\textwidth}
  \includegraphics[width=\textwidth]{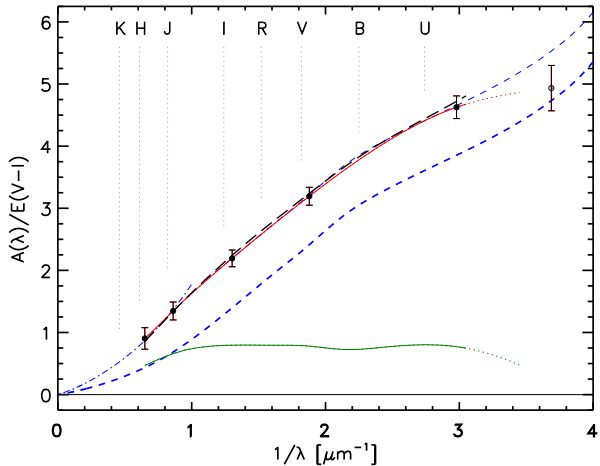}
 \end{minipage}\hfill
 \begin{minipage}[c]{0.3\textwidth}
  \caption{
  The extinction law of the Tarantula nebula (dots and red line) is 
  compared with the one measured inside the nebula's core (R\,136; long-dashed 
  line), and with that of the diffuse Galactic ISM (thick short-dashed 
  line). In the optical domain, the Tarantula and Galactic curves are
  parallel (the green line is their difference). See the text for details.} 
  \label{fig2}
\end{minipage}
\end{figure}

In the Tarantula nebula the reddening vector is much steeper than in the
Galaxy. While in the diffuse Galactic ISM $R_V=A_V/E(B-V)=3.1$ (thick blue dashed 
line in Fig.\,\ref{fig2}), the Tarantula extinction curve (dots and 
red line) implies $R_V=4.5 \pm 0.2$. This value is in excellent agreement 
with the measurements made inside its core R\,136 both from photometry 
($R_V=4.4 \pm 0.2$, black dashed line; De Marchi \& Panagia 2014) and 
spectroscopy ($R_V=4.4 \pm0.4$; Ma\'{\i}z Apell\'aniz et al. 2014). 
At optical wavelengths the extinction law is practically parallel to the 
Galactic law, being shifted to higher values of $R$ by an amount of $0.8$ 
(thin short-dashed line). The difference between the extinction law in 30
Dor and in the Galaxy (green line) is remarkably flat in the optical. 
This indicates that the extinction curve in the Tarantula is due to dust 
of the same type as in the Galaxy, but with an extra component that does 
not depend on wavelength (``grey''). At wavelengths longer than 
$\sim 1\,\mu$m the contribution of this grey component tapers off as 
$\lambda^{-1.5}$, like in the Milky Way (dot-dashed line), further 
suggesting that the nature of the grains is otherwise similar to those in
our Galaxy, but with a $\sim 2$ times higher fraction of large grains. 
This is consistent with the addition of ``fresh'' large grains by type II 
supernova explosions, as recently revealed by the {\em Herschel} and {\em
ALMA} observations of SN\,1987A (e.g. Matsuura et al. 2011; Indebetouw et 
al. 2014). 

These results leave no doubt that in the Tarantula large grains are more 
important than in the diffuse ISM in the Galaxy. These properties are not 
found in the surroundings of the Tarantula, so they must result from 
changes related to the recent/ongoing star formation, implying that the 
ISM is dynamically altered as large fresh grains are produced and later 
destroyed. These findings have important implications for the study of 
the star formation properties in galaxies. Even though these changes 
might last only some $\sim 50$\,Myr, they affect the HII regions at the peak 
of their luminoisty, when they are most easily detectable in distant galaxies. Therefore, 
assuming typical ISM conditions could result in severely underestimated 
total masses and star formation rates, by a factor of 2 or more.

\end{document}